\documentclass[conference]{IEEEtran}
\IEEEoverridecommandlockouts
\usepackage{cite}
\usepackage{amsmath,amssymb,amsfonts}
\usepackage{algorithmic}
\usepackage{graphicx}
\usepackage{textcomp}
\usepackage{xcolor}
\usepackage{listings}
\usepackage{fancyvrb}
\usepackage{framed}
\usepackage[listings,skins]{tcolorbox}
\usepackage[skipbelow=\topskip,skipabove=\topskip]{mdframed}
\mdfsetup{roundcorner=0}
\usepackage{fancyhdr}
\definecolor{dkgreen}{rgb}{0,0.6,0}
\definecolor{altgreen}{rgb}{0.004, .30 , 0.004}
\definecolor{gray}{rgb}{0.5,0.5,0.5}
\definecolor{mauve}{rgb}{0.58,0,0.82}
\definecolor{orange}{rgb}{1.0,0.4,0.05}
\definecolor{mGreen}{rgb}{0,0.6,0}
\definecolor{mGray}{rgb}{0.5,0.5,0.5}
\definecolor{mPurple}{rgb}{0.58,0,0.82}
\definecolor{backgroundColour}{rgb}{0.95,0.95,0.92}
\definecolor{deepblue}{rgb}{0,0,0.5}
\definecolor{deepred}{rgb}{0.6,0,0}
\definecolor{deepgreen}{rgb}{0,0.5,0}\definecolor{dkgreen}{rgb}{0,0.6,0}
\definecolor{altgreen}{rgb}{0.004, .30 , 0.004}
\definecolor{gray}{rgb}{0.5,0.5,0.5}
\definecolor{mauve}{rgb}{0.58,0,0.82}
\definecolor{orange}{rgb}{1.0,0.4,0.05}
\definecolor{mGreen}{rgb}{0,0.6,0}
\definecolor{mGray}{rgb}{0.5,0.5,0.5}
\definecolor{mPurple}{rgb}{0.58,0,0.82}
\definecolor{backgroundColour}{rgb}{0.97,0.97,0.97}
\definecolor{deepblue}{rgb}{0,0,0.5}
\definecolor{deepred}{rgb}{0.6,0,0}
\definecolor{deepgreen}{rgb}{0,0.5,0}
\definecolor{mygreen}{RGB}{28,172,0} % color values Red, Green, Blue
\definecolor{mylilas}{RGB}{170,55,241}

\def\BibTeX{{\rm B\kern-.05em{\sc i\kern-.025em b}\kern-.08em
    T\kern-.1667em\lower.7ex\hbox{E}\kern-.125emX}}

\lstdefinestyle{CStyle}{
    basicstyle=\small\ttfamily,
    backgroundcolor=\color{backgroundColour},   
    keywordstyle=\color{blue},
    commentstyle=\color{mGreen},
    numberstyle=\tiny\color{mGray},
    stringstyle=\color{mPurple},
    breakatwhitespace=false,  
    xleftmargin=1em,
    xrightmargin=1em,
    breaklines=true,                 
    captionpos=b,                    
    keepspaces=true,                 
    numbers=left,                    
    numbersep=5pt,                  
    showspaces=false,                
    showstringspaces=false,
    showtabs=false,                  
    tabsize=2,
    language=C
}

\lstdefinestyle{AStyle}{
    basicstyle=\small\ttfamily,
    backgroundcolor=\color{backgroundColour},   
    commentstyle=\color{mGreen},
    keywordstyle=\color{blue},
    numberstyle=\tiny\color{mGray},
    stringstyle=\color{mPurple},
    breakatwhitespace=false,  
    xleftmargin=1em,
    xrightmargin=1em,
    breaklines=true,                 
    captionpos=b,                    
    keepspaces=true,                 
    numbers=left,                    
    numbersep=5pt,                  
    showspaces=false,                
    showstringspaces=false,
    showtabs=false,                  
    tabsize=2
}

\begin{document}

% \title{An Exploratory Study on Load-Link/Store-Conditional Instructions\\}
\title{Implementing and Breaking Load-Link / Store-Conditional on an ARM-Based System\\}

\author{\IEEEauthorblockN{Evan Tilley}
\IEEEauthorblockA{\textit{Computer Engineering} \\
\textit{Columbia University}\\
New York, United States \\
elt2141@columbia.edu}
\and 
\IEEEauthorblockN{Alexander Liebeskind}
\IEEEauthorblockA{\textit{Computer Engineering} \\
\textit{Columbia University}\\
New York, United States \\
a.liebeskind@columbia.edu}
\and
\IEEEauthorblockN{Rafael Asensio}
\IEEEauthorblockA{\textit{Electrical Engineering} \\
\textit{University College London}\\
London, United Kingdom \\
rafael.asensio.19@ucl.ac.uk}
}
 
\maketitle

\begin{abstract}
Manufacturers of modern electronic devices are constantly attempting to implement additional features into ever-increasingly complex and performance demanding systems. This race has been historically driven by improvements in the processor's clock speed, but as power consumption and real estate concerns in the embedded space pose an growing challenge, multithreading approaches have become more prevalent and relied upon. Synchronization is essential to multithreading systems, as it ensures that threads do not interfere with each others' operations and produce reliable and consistent outputs whilst maximizing performance and efficiency. One of the primary mechanisms guaranteeing synchronization in RISC architectures is the load-link/store conditional routine, which implements an atomic operation that allows a thread to obtain a lock. In this study, we implement, test, and manipulate an LL/SC routine in a multithreading environment using GDB. After examining the routine mechanics, we propose a concise implementation in ARMv7l, as well as demonstrate the importance of register integrity and vulnerabilities that occur when integrity is violated under a limited threat model. This work sheds light on LL/SC operations and related lock routines used for multithreading.
\end{abstract}


\section{Introduction}
%Load-link/store-conditional instructions are \\
%- issue with multiple threads accessing same variable \\
%- solutions: test \& set + LLSC

Modern day computing systems rely heavily on multithreading to increase performance and resource utilization through improved responsiveness, resource sharing, economy and scalability. However, running multiple threads concurrently can become a complex challenge, with individual threads being capable of interfering with each other stochastically when sharing hardware resources. For instance, when multiple threads try to access a shared variable concurrently, the read and write operations can overlap in execution due to race conditions which can lead to unpredictable outputs depending on the order of access.

Thread interference errors can be corrected by ensuring synchronization, which is established by satisfying two conditions:
\begin{enumerate}
    \item Only one thread can access a shared variable at a time, guaranteeing that the variable is atomic and other threads do not interfere with the operation.
    \item Whenever a thread accesses a shared variable, it ensures that the changes are visible to other subsequent threads by establishing a happens-before relationship with ensuing accesses.
\end{enumerate}

One method of addressing synchronization is load-linked/store-conditional (LL/SC), which refers to a pair of instructions that protect multithreaded accesses to memory by implementing a lock-free atomic read-modify-write (RMW) operation. The first instruction, LL or load-link, implements the \textit{read} part of the atomic operation in two stages. First, LL reads from a memory location and puts the read value in a register. Second, the LL instruction appropriately updates values in the architecture (exclusive monitors in ARM systems), based on the address from which the value was loaded. The second instruction, SC or store-conditional, implements the \textit{write} part of the atomic operation. The SC instruction attempts to store a value to an address in memory. If this store is permitted by the system (which is likely in the case that another thread hasn't stored to the address and already obtained the lock), then the store will succeed (a 0 will be returned in the destination register in ARM). Otherwise, the store will fail (a 1 will be returned in the destination register in ARM). As such, the store is known as a conditional store, which is not guaranteed to succeed. The LL and SC operations work together with exclusive monitors in ARM architectures to provide atomic memory updates.

It is worth mentioning that LL/SC instructions are only compatible with RISC architectures such as ARM or MIPS. CISC architectures, like x86, implement a similar instruction called compare-and-swap (CAS), which has a slightly different functionality. Though implementation methods for LL/SC vary, some rely upon CAS \cite{b1}. Specific syntax for LL/SC instructions varies across platforms. 

Given the prevalence of LL/SC instructions, their functionality and potential vulnerabilities are of significant interest. This study aimed to illuminate LL/SC by implementing a lock scheme in ARMv7l using a Raspberry Pi through the supported \texttt{LDREX} and \texttt{STREX} instructions. We also subsequently explored the operation of these operations when executed by several threads and attempted to introduce unpredicted malicious behavior.

This paper overall makes the following contributions:
\begin{itemize}
\item Concise implementation of LL/SC in a multithreading environment on an ARM-based system
\item Demonstration of the importance of register integrity during the LL/SC routine by highlighting the the impact an attacker can have by tampering with bits in key registers.
\item Several implementation suggestions that may help protect register integrity
\end{itemize}

\section{Key Idea}
Synchronization is essential to various real-world use cases. Consider the example of a bank account's details stored in memory, in which an \texttt{accountBalance} variable stores a numeric value for a given account (Fig. \ref{code_1}). Naturally, a synchronization failure could cause an error (or even a security breach), given that calling \texttt{increaseBalance()} as shown in Fig. \ref{code_1} without appropriately-placed thread barriers may generate unexpected behavior which could be exploited by a malicious user. If two threads (access points) attempt to increment the account balance at the same time, they may each fetch the initial value and add to it, instead of waiting for the other thread to finish its execution and then adding to the resulting updated value. This means that the thread that concludes last will have effectively called \texttt{increaseBalance()}, but with no noticeable effect. Similar programs are commonly used to demonstrate the danger of sharing a global variable between threads.

\begin{figure}[hbtp]
    \begin{lstlisting}[style=CStyle,frame=single]
double accountBalance = 0;
void *increaseBalance(){
    int i;
    for (i = 0; i < 10000; i++){
        accountBalance += 1;
    }
} \end{lstlisting}
    \caption{Simple code for increasing an account balance through incrementation}
    \label{code_1}
\end{figure}

%More concretely, we called the code in  Fig. \ref{code_1} which gave the following rresults...
%For instance, if 3 threads callfunction (without barriers in place), the result will not be 30,000 as expected. Running this with 3 threads, we observed results ranging from \textbf{GREAT PLACE FOR DATA, INCLUDE GRAPH HERE} x to y.
%[see appendix here -- full code included]

This idea of unexpected multithreading behavior serves as a source of motivation for a mechanism ensuring that only one thread is accessing a shared variable at a time. It is not difficult to imagine circumstances in which multiple threads executing privileged code in an uncoordinated manner could lead to catastrophe by modifying a critical variable in an unexpected way. It is also easy to conceive of why privileged code should not be accessed at the same time. Since tangible issues could arise from synchronization bugs, further research advancing knowledge on LL/SC instructions and lock sequences could be influential.

% \section{Key Idea}
% T

% \section{Key Idea}
% The motivation for this study was that LL/SC instructions may potentially allow for a vulnerability, which might be exploited through a side-channel or other means to reveal information about the registers being accessed or the stored values. As a result, we aimed to manipulate and explore both the LL/SC lock routine itself and the multithreading approach used to try and obtain locks and execute privileged code.

\section{Methodology}

\subsection{Overview}
We aimed to create a program capable of implementing and testing LL/SC in a multithread environment. This was achieved using ARMv7l on a Raspberry Pi through a C harness file with inline ASM code. In order to ensure synchronization of variables between different threads, we utilized GDB to step across multiple threads and ensure that the routine was working as expected. Finally, we attempted to break the LL/SC implementation by manipulating register values at runtime, and analyzing the results and their implications on the routine.

\subsection{LL/SC Implementation}
For the base LL/SC implementation, we relied upon supported inline ASM commands to create the lock routine. More specifically, our locking scheme utilized the \texttt{LDREX} and \texttt{STREX} instructions in combination with the \texttt{BNE} command to jump to the start of the program if a compare (\texttt{CMP}) operation failed. We kept track of the shared lock variable in the C harness code through \texttt{lockVar}.
\begin{figure}[hbtp]
    \begin{lstlisting}[style=AStyle,frame=single]
lock:
    retry:
        LDR R10, =lockVar
        LDREX R8, [R10]
        CMP R8, #0
        BNE retry
        MOV R9, #1
        STREX R2, R9, [R10]
        CMP R2, #0
        BNE retry

    critical_section:
        // critical shared variable
        // changes could occur here
    
    unlock:
        MOV R5, #0
        LDR R10, =lockVar
        STR R5, [R10] \end{lstlisting}
    \caption{Lock routine utilizing \texttt{LDREX}, \texttt{STREX}, and other built in operations for LL/SC in inline ASM}
    \label{code_2}
\end{figure}

Our ASM code was modeled after prevailing LL/SC schemes in computer architecture. In the \texttt{retry} routine, a thread attempts to write to a lock. If the value loaded in through the \texttt{LDREX} instruction is 0 (i.e., no thread currently has the lock at the time the instruction was executed), the thread then attempts to write a value of 1 to the shared lock variable in the \texttt{STREX} instruction. If successful, the thread will then move on to the critical section.

After the critical section, the thread will release its hold on the lock by writing a value of 0 to \texttt{lockVar}, at which points other threads spinning in the \texttt{retry} routine can gain hold of the lock. The \texttt{CMP} \texttt{BNE} pair recursively calls \texttt{retry}, ensuring that a thread will repeatedly try to gain the lock if it is already taken. So long as there is are competing threads, one thread giving up the lock will almost immediately yield it to another thread. \texttt{LDREX} and \texttt{STREX} rely on exclusive monitors, as specified by the ARM architecture.

\subsection{Multithreading}
Our ASM LL/SC routine was implemented with inline ASM in C, which allowed us to build out a more robust testing platform. The ASM code was wrapped in a void \texttt{*loadLinkRoutine} function, which was then called by threads from a main function within the C program. The main function also handled creating threads using the pthread library, allowing us to index threads and monitor access patterns, which was critical to keep track track of "winning" and "losing" threads. A simple loop handled thread generation, so that it was possible to vary the number of threads throughout experimentation and observe resulting trends in race conditions and thread execution order.

\begin{figure}[hbtp]
    \begin{lstlisting}[style=CStyle,frame=single]
int main(){
    int i;
    pthread_t tid;
    for (i = 0; i < 10; i++){
        pthread_create(&tid, NULL,
        loadLinkRoutine, NULL);
    }
    pthread_exit(NULL);
    return 0;
} \end{lstlisting}
    \caption{C code calling the lock routine and handling multithreading}
    \label{code_3}
\end{figure}

% Note that there are no barriers to control thread execution and so slight variations in thread start times will create race conditions, which creates an ideal environment for analyzing the behavior of our LLSC routine.
% It is fundamental to note that looping alone does not account for thread timing. Solely with a loop and no barriers to control thread execution, slight variations in thread start times due to loop timing (the time delay between loops mean that earlier threads have a head start) result in sequential thread operation. In other words, the threads would execute in the order in which they were created, preventing a race condition. As a result, it is necessary to leverage \texttt{pthread\_barrier} to make sure that all threads start at exactly the same time. This simulates threads accessing the lock at the exact same moment. 

\subsection{Testing Challenges}

The combination of C code and inline ASM proved to be a challenge when attempting to test the LL/SC routine. In particular it was difficult to directly observe outputs because even if the ASM routines are split up into several C functions which are chained together (allowing inter-routine \texttt{printf} calls to be execution), it is not possible to guarantee the order of the \texttt{printf} calls, as the threads are racing one another.

An alternative method of debugging would be to insert the print statements \textit{directly into} the ASM code. For instance, it is possible to take advantage of extended ASM to insert print statements as shown in Fig. 4.

\begin{figure}[hbtp]
    \begin{lstlisting}[style=CStyle,frame=single]
#include <string.h>
int main() {
    char* str = "Hello World\n";
    long len = strlen(str);

    __asm__(
    "MOV R0, #2\n\t" // stderr
    "MOV R1, %[toPrint]\n\t" // message
    "MOV R2, %[length]\n\t" // length
    "MOV R7, #4\n\t"
    "SWI 0"
    :
    : [toPrint] "r" (str),
      [length] "r" (len)
);
    return 0;
} \end{lstlisting}
    \caption{Direct print statements from ASM}
    \label{code_4}
\end{figure}

Unfortunately, this approach was not effective given that if multiple threads cause a software interrupt at the same time, the output of one thread will overwrite the other. In certain scenarios, before the full string output of one thread can be printed out, another thread may make a software interrupt and prevent the former call from being fully executed. Without artificially guaranteeing the order of threads using barriers (and defeating the purpose of having multiple threads simultaneously accessing the same shared variable) a consistent output cannot be assured.

\begin{figure}[htp]
    \centering
    \includegraphics[width=9cm]{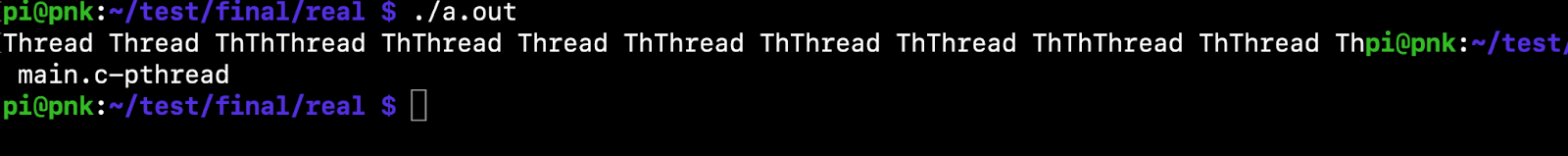}
    \caption{Unconstrained behavior with incorrect threading implementation}
    \label{fig:galaxy}
\end{figure}

More advanced capabilities are required for lower level access, necessitating use of GDB to test the functionality of the routine in a more rigorous manner. This solution met the demands of our research objective while allowing us to retain an ASM and C code framework. 

\subsection{GNU Debugger (GDB)}

GNU Debugger, or GDB, is a portable debugger that is typically run from Unix-like systems. It grants access to registers and memory, and allows the programmer to insert breakpoints, pass commands at runtime, and inspect the functionality of code at a low level. GDB hence allowed us to probe and test the routine more precisely than print statements or other discussed modifications to the code. In order to understand and explore the behaviour of multiple threads racing against each other, we set a breakpoint at the start of the \texttt{*loadLinkRoutine} function so that each thread will stop execution here and allow us to step through the following instructions one-by-one.

GDB cannot single-step all threads in lockstep \cite{b2}, making analyzing the behaviour of multiple threads racing against each other difficult because, by default, other threads can execute more than one statement whilst the current thread only completes a single step. The Raspian OS, however, does support locking the OS scheduler so that a single thread can run at a time. This behavior can be achieved through GDB by executing the command: 

\texttt{set scheduler-locking step}

This scheduler locking mechanism allowed us to manually stop each thread and have direct access to threads attempting to modify the same variable at once.

\subsection{Experimentation Model}
Early versions of our LL/SC implementation and attack model did not produce the desired behavior, driving us to improve upon the lock routine and develop new methods of thread exploitation. We debugged and validated the functionality of LL/SC using GDB as described to ensure that the attack was independent of our implementation. 

Prior to recording results and engineering our attack, we also dedicated a significant research effort towards LL/SC lock design. This experimentation involved introducing additional branch conditions (with \texttt{BNE}), \texttt{LDREX} without \texttt{STREX} (and vice versa), and misordered register accesses. The goal of this iterated testing was to fully determine LL/SC mechanics outside of traditional use before moving on to results. 

The majority of errors we encountered with this experimentation were trivial and did not reveal any vulnerabilities related to LL/SC. Given the syntax and register usage of \texttt{LDREX} and \texttt{STREX}, there are limitations to the manner in which these instructions can viably be used. In some cases we reached segmentation faults and bus errors when LL/SC was improperly implemented. Still, this experimentation helped to narrow the scope of our attack and verify ways in which LL/SC was not prone to vulnerability.  

%\textbf{COMPLETE THIS}
%ASM
%C Threading w pthread
%Testing
%- gdb (insert images)
%link "Debugging with GDB" (webpage)
%synced threads to same step
%thread by thread showing we have the lock
%- loadlinkRoutine
%Modifications

\section{Results}
\subsection{Lock Sequence}
In order to ensure the functionality of our LL/SC routine, we followed a three part process:
\begin{enumerate}
    \item Step the winner thread to the point where it had obtained the lock on the critical variable.
    \item Step the remaining threads (the losers) to the 
    \texttt{LDREX} instruction, and monitor their behavior.
    \item Release the winner's hold on the lock. Repeat.
\end{enumerate} 
This process verified that our LL/SC code was working properly. The winning thread maintained the lock when expected and executed the \texttt{critical\_section}, while the losing threads looped in attempts to gain the lock. This was repeated for subsequent threads as described above.

While validating our lock sequence, we also used GDB to verify thread "locations" relative to the lock routine and register values as threads executed. This provided a low level view into memory without relying on the ASM or C code to examine program behavior. 

\subsection{Observations}
After the LL/SC implementation was shown to be a functional lock routine, and prior to attempting to alter the lock or break LL/SC, we first conducted preliminary testing to determine areas of vulnerability for exploit. This allowed us to fine tune our attack on LL/SC based on thread execution. One interesting discovery we made during this phase was that while the winning thread had a hold on the lock, the losing threads would continually cycle between the instruction pair shown in Fig. \ref{code_5}.

\begin{figure}[hbtp]
    \begin{lstlisting}[style=CStyle,frame=single]
LDR R10, =lockVar
LDREX R8, [R10] \end{lstlisting}
    \caption{Instruction pair which losing threads cycled over after the lock was obtained by the winning thread}
    \label{code_5}
\end{figure}

This was an interesting result because it meant that we could not directly step into the \texttt{CMP} statement shown in Fig. \ref{code_6}, due to the fact that the \texttt{LDREX} and \texttt{STREX} instructions are treated as atomic by ARM. Stepping between the \texttt{LDREX} and \texttt{STREX} instructions would essentially violate the atomicity of \texttt{LDREX} and \texttt{STREX}, breaking the pair into independent instructions. 

\begin{figure}[hbtp]
    \begin{lstlisting}[style=CStyle,frame=single]
CMP R8, #0 \end{lstlisting}
    \caption{\texttt{CMP} statement that could not be reached by GDB due to atomic \texttt{LDREX} and \texttt{STREX} instructions}
    \label{code_6}
\end{figure}

To prevent this, GDB does not allow one to step through the instructions in between LDREX and STREX which proved to have important consequences when examining security vulnerabilities.

\subsection{Breaking LL/SC}
In order to attempt to break our implementation of LL/SC, and hence showcase the importance of LL/SC in synchronizing threads, we created numerous threads to attempt to access a memory address at the same time. More specifically,  we created 3 threads, each of which added 5 to an initial \texttt{accountBalance} of 100 (representing a bank account). To ensure that the resulting \texttt{accountBalance} could be correctly printed when the program terminated, we utilized \texttt{pthread\_barrier}. 

Given that each of the 3 threads added 5 to an initial balance of 100, the expected output of the program was 115. However, by modifying register values during runtime, we were able to break the locking mechanism and cause two threads to access the \texttt{accountBalance} variable at the same time, inducing the output to vary from expectation.

As we found in our observation period, it was not feasible to modify instructions that take place between the \texttt{LDREX} and \texttt{STREX}, due to the atomicity of this operation pair. Instead, we needed to modify the LL/SC routine slightly so that the the CMP instruction compares two registers, as opposed to one register (the result of the \texttt{LDREX}/\texttt{STREX}) and a constant. This step would crucially allow us to modify the second register during runtime through the modified routine found in Fig. \ref{code_7}. 

\begin{figure}[hbtp]
    \begin{lstlisting}[style=CStyle,frame=single]
lock:
    retry:
        LDR R10, =lockVar
        MOV R7, #0
        LDREX R8, [R10]
        CMP R8, R7 // updated comparison
        BNE retry
        MOV R9, #1
        STREX R2, R9, [R10]
        CMP R2, R7 // updated comparison
        BNE retry

    critical_section:
        // critical shared variable changes would occur here
    
    unlock:
        MOV R5, #0
        LDR R10, =lockVar
        STR R5, [R10] \end{lstlisting}
    \caption{Modified lock routine}
    \label{code_7}
\end{figure}

With this updated routine, we were able to successfully modify the value in \texttt{R7} at runtime and produce an outcome that deviated from the expected 115 through pathological register manipulation and thread stepping.

To begin, we ran GDB using the described methodology, taking advantage of the step scheduler-locking mechanism to step through threads in place. Our three threads each started at the beginning of the \texttt{loadLinkRoutine} function. Utilizing GDB, Thread 1 first obtained the lock and added the value of five to \texttt{accountBalance}. Next, \textit{before Thread 1 had stored the value of 0 back to the lock (i.e., freed the lock)}, we switched to Thread 2. As expected, this thread was continually cycling between the three instructions detailed in Fig. \ref{code_8}.

\begin{figure}[hbtp]
    \begin{lstlisting}[style=CStyle,frame=single]
LDR R10, =lockVar
MOV R7, #0
LDREX R8, [R10] // updated comparison \end{lstlisting}
    \caption{Cycled instructions for Thread 2}
    \label{code_8}
\end{figure}

We then used GDB to change the value in the register \texttt{R7} immediately after the \texttt{MOV} instruction. Since we could not modify \texttt{R7} at any point in between the \texttt{LDREX} and \texttt{STREX}, we modified it before the \texttt{LDREX} instruction executed. 

The instruction \texttt{set \$R7 += 1} was executed in GDB so that the comparison between \texttt{R8} and \texttt{R7} (which should return 0 under normal circumstances since Thread 1 has the lock and therefore the result of \texttt{LDREX}, stored in \texttt{R8}, will be 1) returned 1. This manipulation caused Thread 2 to make it past the \texttt{STREX} instruction. One instruction, however, still laid in the way: the \texttt{CMP} after \texttt{STREX}. This instruction compared \texttt{R2} and \texttt{R7}. As a result, we needed to manipulate \texttt{R7} once more for the line \texttt{CMP R2, R7}. This time, it was necessary to set \texttt{R7} to 0, which would allow us the thread to move forward to the next instruction instead of continually spinning for the lock. This modification permitted Thread 2 to make it past the second \texttt{CMP} instruction, on the assumption that it now had the lock. Thread 1 also thought it had the lock. 

It is deeply problematic that Thread 1 and Thread 2 simultaneously believed they each had the lock, as this allowed Thread 2 to read in the value of the \texttt{}{accountBalance} prior to update by Thread 1. Thread 2 read in an \texttt{accountBalance} value of 100, when it really should have read in a value of 105 (after it was updated by Thread 1). This meant that the base value of 100 was updated, not 105. After causing this crucial break in our LL/SC subroutine, we resumed normal operation by turning off the step mechanism via:

\texttt{set scheduler-locking off}.

After all threads finished running, we received the output shown in Fig. \ref{code_9}.  
\begin{figure}[hbtp]
    \begin{lstlisting}[style=CStyle,frame=single]
after all threads have run, the value of account balance is 110
[Thread 0x75e3e460 (LWP 2152) exited]
[Thread 0x7663f460 (LWP 2151) exited]
[Thread 0x76e40460 (LWP 2150) exited]
[Inferior 1 (process 2147) exited normally]
\end{lstlisting}
    \caption{Thread output at conclusion of runtime}
    \label{code_9}
\end{figure}

This output of 110 instead of the expected 115 illustrated that it was possible to break our LL/SC routine and cause an unexpected result through runtime register manipulation. 

\subsection{Implementation Suggestions}
Based on our results, we suggest the following two measures to increase security in LLSC implementations:
\begin{enumerate}
    \item Ensure that the outputs of the LL and SC instructions are compared to constants, and not registers.
    \item Have a branch conditional both after the LL instruction and after the SC instruction.
\end{enumerate}

The first recommendation is a direct result of the behavior observed through pathological register manipulation where we were able to modify the \texttt{accountBalance} variable in a manner different than what was presumably intended by the program(mer). The actual output of the LDREX and STREX instructions can not be modified in between the LDREX and STREX instructions due to the atomic nature of LDREX and STREX working in conjunction. So, for instance, the code in Fig. 11 is safe because the \texttt{R8} register cannot be modified, and neither can the constant value.
\begin{figure}[hbtp]
    \begin{lstlisting}[style=CStyle,frame=single]
    LDREX R8, [R10]
    CMP R8, #0
\end{lstlisting}
    \caption{Thread output at conclusion of runtime}
    \label{code_11}
\end{figure}

The code in Fig. 12, however, is potentially unsafe because \texttt{R7} can be modified before the LLSC routine occurs, as we demonstrated.

\begin{figure}[hbtp]
    \begin{lstlisting}[style=CStyle,frame=single]
    LDREX R8, [R10]
    CMP R8, R7
\end{lstlisting}
    \caption{Thread output at conclusion of runtime}
    \label{code_12}
\end{figure}

The second recommendation relates to the number of modifications that must be made to corrupt the LLSC routine. Some implementations do not include a branch based on the result of LDREX, such as the implementation depicted in Figure 13. 

\begin{figure}[hbtp]
    \begin{lstlisting}[style=AStyle,frame=single]
lock:
    retry:
        LDR R10, =lockVar
        LDREX R8, [R10]
        MOV R9, #1
        STREX R2, R9, [R10]
        CMP R2, #0
        BNE retry

    critical_section:
        // critical shared variable
        // changes could occur here
    
    unlock:
        MOV R5, #0
        LDR R10, =lockVar
        STR R5, [R10] \end{lstlisting}
    \caption{Lock routine without branch depending on output of \texttt{LDREX}}
    \label{code_13}
\end{figure}

Thus, the only instruction standing in the way of LLSC corruption is the sole \texttt{CMP} instruction after \texttt{STREX}, as compared to our initial implementation with had a \texttt{CMP} instruction both after \texttt{LDREX} and \texttt{STREX}.

\section{Related Work}
\subsection{Implementation and Optimization}
A number of prior studies have explored the mechanics of LL/SC instructions, compared LL/SC to alternatives such as CAS, and explored the implications of LL/SC support across platforms. Notably, Alpha, PowerPC, MIPS, and ARM all provide LL/SC instructions, while x86 leverages CAS \cite{b3}. This means that depending on the hardware and CPU, LL/SC is sometimes infeasible. In our study, for instance, we initially approached x86 before opting for ARM. LL/SC differs from CAS in that it is more rigorous: an successful LL/SC programming pattern assures that the store-conditional will fail if any updates have occurred to the memory in question, while CAS solely verifies that the value is maintained. In practice, LL/SC guest instructions are often produced from host atomic CAS instructions, as is necessary in the case of x86 host systems \cite{b1, b4}. Though efficient, this has been shown to induce significant issues, which requires a new LL/SC emulation scheme \cite{b1}. Research has also investigated improving upon the implementation of CAS \cite{b5}. 

In the context of this work, our exploration of LL/SC routines and attacks helps to emphasize areas for future research. In implementing and optimizing LL/SC, it is critical that engineers bear security in mind, as synchronization could fail without the proper lock routine requirements. Improving upon existing LL/SC schemes must not sacrifice security for performance. Additionally, it is notable that lock register integrity remains incredibly important in terms of ensuring functional LL/SC instructions. Memory constraints (and which/how many registers are used in LL/SC implementations) must be prioritized.

\subsection{LL/SC Usage}
In addition to research investigating LL/SC directly, a large effort has been devoted to examining LL/SC in the context of related computing research. Given the pertinence of LL/SC operations in RMWs and the existence of supported open source standards (e.g. RISC-V), LL/SC is widely useful on RISC machines \cite{b6}. For instance, LL/SC operations are pivotal in generating atomic instructions in RISC machines for emulating multithreaded applications on multicore systems \cite{b7}. 
Instruction set extensions introduced in research sometimes share properties with LL/SC \cite{b8}, meaning that the pattern of memory access is often similar. In such circumstances, LL/SC is conceptually linked as a blueprint for instruction dependency and memory maintenance. Novel primitive operations that are generalizations of the \texttt{LDREX} and \texttt{STREX} instructions have also been proposed for specific uses such as non-blocking data structures, further demonstrating the deployment of LL/SC-like instruction sequences \cite{b9}. It is possible that our work applies to LL/SC-like instruction routines, though this would likely depend on the mechanics of alternative operation pairs. In any case, our work is of interest to software and hardware researchers in fields wherein LL/SC is common, even if the immediate topic of interest is not lock routines.

\section{Conclusion}

This paper examined a concise implementation of LL/SC on ARMv7l, demonstrated the importance of register integrity in a multithreading environment, and highlighted a few important security considerations to take into account when implementine LL/SC.

We built a C program to test the LL/SC routine under multithreading conditions, supervising its functioning using GDB and stepping through the instructions line by line. Based on our experimentation and results, we concluded that LL/SC can be exploited by an attacker with access to key individual registers, breaking synchronization and opening the door for vulnerabilities. Furthermore, we think that some of the assumptions of the threat model can be bypassed by combining our results with previously developed attacks. For instance, attacks like Row Hammer \cite{b10} are capable of flipping bits in registers, allowing for an attacker to modify the key register value that allows our exploit to succeed. Cloud computing companies may place increasing emphasis on the integrity of on-chip registers.

This finding demonstrates a broader principle for lock routines: a lock is only as strong as the registers storing essential values. LL/SC, for instance, relies on several aforementioned registers called during the lock routine. By modifying these registers, a malicious attacker might be able to undermine synchronization.

We hope that our findings serve as a stepping stone for further research into exploiting the inbuilt assumptions that go into the atomicity of the LL/SC routine, and that it can be combined with other attacks to strengthen the capabilities and versatility of the load-link/store-conditional mechanism.

%  to establish a more complete threat model to strengthen the capabilities and versatility of the exploit to gather useful data via side-channels.

\section*{Acknowledgment}
We would like to thank Professor Sethumadhavan for his dedicated guidance on this research. We would also like to recognize the undergraduate and graduate students in Hardware Security at Columbia University for a topical discussion of hardware vulnerabilities, side-channels, and attacks on memory.

\section*{Appendix}
While this appendix is not integral to understanding of our research, additional code is provided to facilitate replicability and make our approach more clear. The code that follows was used throughout the process of implementation, experimentation, and attacking LL/SC. 

\subsection*{Code Repository}
Our full code can be found on Github, including our final code and significant portions of experimentation, at the following repository:\\

https://github.com/evantilley/hardware\_sec

\subsection*{Code Description}
A brief video explaining our code can be found at the following link:\\

https://www.youtube.com/watch?v=b1YQlx8vFc4

\begin{figure}[hbtp]
\subsection*{\texttt{main.c}:}
    \begin{lstlisting}[style=CStyle,frame=single]
#include <stdio.h>
#include <string.h>
#include <stdlib.h>
#include <pthread.h>

int lockVar = 0;

void *loadLinkRoutine(){
    long threadID = (long) pthread_self();

    __asm__ volatile(
    "lock:\n\t"
    "try_again:\n\t"
    // "MOV R3, %[lock]\n\t" // move lock variable into R3
    "LDR R10, =lockVar\n\t"
    "LDREX R8, [R10]\n\t" // load value stored at that address
    "CMP R8, #0\n\t"
    "BNE try_again\n\t"
    "MOV R9, #1\n\t"
    "STREX R2, R9, [R10]\n\t"
    "CMP R2, #0\n\t"
    "BNE try_again\n\t"
    // if we make it here, got the lock
    "MOV R5, #0\n\t"
    "LDR R10, =lockVar\n\t"
    "STR R5, [R10]\n\t"
    //"LDREX R3, =lockVar" // [lock must be an address]
    ::
      [lock] "r" (lockVar)
);
} \end{lstlisting}
    \caption{Beginning of \texttt{main.c}, the full normal functioning LL/SC}
    \label{code_10}
\end{figure}

\begin{figure}[hbtp]
    \begin{lstlisting}[style=CStyle,frame=single]
int main(){
    int i;
    pthread_t tid;
    for (i = 0; i < 10; i++){
         pthread_create(&tid, NULL, loadLinkRoutine, NULL);
    }
    pthread_exit(NULL);
    return 0;
} \end{lstlisting}
    \caption{Main function of \texttt{main.c}, the full normal functioning LL/SC}
    \label{code_11}
\end{figure}

\begin{figure}[hbtp]
\subsection*{\texttt{hacker.c}:}
    \begin{lstlisting}[style=CStyle,frame=single]
#include <stdio.h>
#include <string.h>
#include <stdlib.h>
#include <pthread.h>

int lockVar = 0;
int accountBalance = 100;
pthread_barrier_t barrier;

void *loadLinkRoutine(){
    long threadID = (long) pthread_self();
    int realBal = accountBalance;
     __asm__ volatile(
    "lock:\n\t"
    "try_again:\n\t"
   // "MOV R3, %[lock]\n\t" // move lock variable into R3
    "LDR R10, =lockVar\n\t"
    "MOV R7, #0\n\t"
    "LDREX R8, [R10]\n\t" // load value stored at that address
    "NOP \n\t"
    "CMP R8, R7\n\t"
    "NOP \n\t"
    "BNE try_again\n\t"
    "MOV R9, #1\n\t"
    "STREX R2, R9, [R10]\n\t"
    "CMP R2, R7\n\t"
    "BNE try_again\n\t"
    // if we make it here, got the lock
    // critical section - modify shared variable here
    "MOV R4, #5\n\t"
    "add %[DEST], R4\n\t"

    // okay, shared variable has been modified
    // now, replace the lock
    "MOV R5, #0\n\t"
    "LDR R10, =lockVar\n\t"
    "STR R5, [R10]\n\t"
       
    : [DEST] "=r" (accountBalance)
    //  [balRes] "=r" (accountBalance)
    : "[DEST]" (accountBalance)
);
    pthread_barrier_wait(&barrier);
} \end{lstlisting}
    \caption{Beginning of \texttt{hacker.c}, the modified code that theoretically outputs an expected value of 115, which we successfully tampered with using GDB on registers to reach an output of 110}
    \label{code_12}
\end{figure}
\begin{figure}[hbtp]
    \begin{lstlisting}[style=CStyle,frame=single]
int main(){
    int i;
    pthread_t t[3];
    pthread_barrier_init(&barrier, NULL, 4);

    pthread_create(&t[0], NULL, loadLinkRoutine, NULL);
    pthread_create(&t[1], NULL, loadLinkRoutine, NULL);
    pthread_create(&t[2], NULL, loadLinkRoutine, NULL);

    pthread_barrier_wait(&barrier);
    fprintf(stderr, "after all threads have run, the value of account balance is %d\n", acco$
    pthread_barrier_destroy(&barrier);
    return 0;
} \end{lstlisting}
    \caption{Beginning of \texttt{hacker.c}, the modified code that theoretically outputs an expected value of 115, which we successfully tampered with using GDB on registers to reach an output of 110}
    \label{code_13}
\end{figure}

\end{document}